\documentclass[useAMS,usenatbib]{mnras}

\usepackage{color}
\usepackage[english]{babel}
\usepackage{graphicx}
\usepackage{amsmath}
\usepackage{amssymb}
\usepackage{epsf}
\usepackage{bm}
\usepackage[utf8]{inputenc}

\begin{document}

% *************************************************************************
%                                 TITLE
% *************************************************************************

%\begin{document}

\title[Breaking properties of multicomponent neutron star crust]{Breaking properties of multicomponent neutron star crust}

\author[A. A. Kozhberov]{{A. A. Kozhberov\thanks{E-mail: kozhberov@gmail.com}}\\
    Ioffe Institute, Politekhnicheskaya 26, St~Petersburg 194021, Russia\\
    }

%\date{Accepted . Received ; in original form}
%\pagerange{\pageref{firstpage}--\pageref{lastpage}} \pubyear{2010}

\maketitle \label{firstpage}

\begin{abstract} 
We study breaking properties of a solid neutron star crust. We consider the case in which the crust at any fixed density consists of two types ions, forming a strongly ordered Coulomb crystal. It is shown that the breaking stress of a such matter noticeably depends on ionic composition, and it is typically larger than for a one-component crystal. The difference may reach a factor of several.
\end{abstract}


\section{Introduction}
The crust of a neutron star is mostly composed of atomic nuclei, degenerate electrons and possibly free neutrons \citep[e.g.,][]{ST1983,HPY2007}. Although the crustal mass is about $1$ per cent of the total stellar mass, the crust is important for evolution and different observational manifestations of neutron stars \citep[e.g.,][]{CH08}. In this work, we continue our investigations of the behavior of the solid crust under deformations. Our previous article \citep{K19} was mostly devoted to the effective shear modulus, but the current paper focuses on the breaking properties.

The maximum breaking stress ($\sigma_{\max}$) is important in many aspects of the neutron star theory. It determines the maximum quadrupole moment and ellipticity of the star \citep[e.g.,][]{UCB00,O05,JO13}, which in turn can be crucial for gravitational wave observations of neutron stars \citep[e.g.,][]{JKS98,A19,KM22}. $\sigma_{\max}$ is used in studying magnetars, in solving global problems of the magneto-thermal evolution of neutron stars \citep[e.g.,][]{PP11,VRP13,THN17,BJA20,GTWZ20} and, in particular, the problems associated with mechanical failures such as plastic motions, starquakes, outbursts \citep[e.g.,][]{BL14,LAAW15,LG19,LL16,L16,KB17}. Also, the breaking stress is applied for investigating oscillations of neutron stars \citep[e.g.,][]{FHL18,KY20,PAP21,KSK21} and pulsar glitches \citep[see][and reference their in]{PFH14,RAR21,ZG22}.

The modern models suggest that the accreting neutron star crust is composed of a mixture of various ions \citep[e.g.,][]{DG09,HB09,SC19}. The solid neutron star crust is usually described by the model of a Coulomb crystal of point-like ions \citep[e.g.,][]{HPY2007}. Hence it is instructive to consider breaking properties of multi-component Coulomb crystals. 

Contemporary view on the breaking stress of the neutron star crust is based on molecular dynamic simulations of ions interacting via the screened Coulomb potential. They were performed by \cite{HK09,CH10,CH12,HH12}. Later the problem was investigated analytically by \cite{BK17,BC18} for the body-centered cubic (bcc) lattice with the uniform electron background, where a significant dependence $\sigma_{\max}$ on the direction of deformation was found. In \cite{KY20} the Coulomb crystals considered in details in the same way as by \cite{BK17}. It was shown that effects of electron polarization (plasma screening) could be neglected. 

Some multi-component ionic crystals were already considered using numerical simulations by \cite{HK09,HH12}, but breaking properties of strongly ordered binary crystals have not been explored before current work. In this paper we study two directions of deformation. We performed semi-analytic calculations of the critical strain and breaking stress for these directions as a function of the ratio of ion charges in the lattice.

\section{Critical strain and breaking stress in binary Coulomb crystals}
Current models of the internal structure of neutron stars predict that at densities 
$\rho \sim 10^6-10^{11} {\rm g~cm^{-3}}$ their envelopes contain only atomic nuclei and strongly degenerate electrons. The question of variation of atomic nuclei parameters and properties with depth is complicated and depends on many factors (e.g., \citealt{HPY2007}), such as a model of the equation of state and a history of stellar evolution (accretion, flares). 

If all ions have identical charge number $Z$ and interact via Coulomb forces, the ions are crystallized at coupling parameter $\Gamma \gtrsim 175$, where
\begin{equation}
   \Gamma \equiv \frac{Z^2 e^2}{ak_{\rm B} T}~, 
\end{equation}	 
$a \equiv \left({4 \pi n}/3 \right)^{-1/3}$ is the ion-sphere radius, $T$ is the temperature, $e$ is the elementary charge, $k_{\rm B}$ is the Boltzmann constant, $n=N/V$ is the number density of the ions, $N$ is the number of ions, $V$ is the volume of the system.

Phase diagrams of binary (with two types of ions having charge numbers $Z_1$ and $Z_2$) ionic mixtures are more complicated (e.g., \citealt{B22}). However we can assume that at low temperatures ionic mixtures solidify usually to an ordered binary bcc lattice, which can be described by the binary Coulomb crystal model (e.g., \citealt{K20}). 

For the first time, the breaking stress ($\sigma_{\max}$) of solid neutron star crust was obtain from molecular dynamic simulations by \cite{CH10,CH12}. In their model all ions have the same charge number, form a perfect bcc lattice and interact via a screened Coulomb potential, while $\sigma_{\rm max}$ was determined through direct molecular dynamic simulations of crystal evolution under increasing deformation. \cite{CH10} considered only one direction of deformation. It translates the vector $a_{\rm l}(n_1,n_2,n_3)$ as
\begin{equation}
a_{\rm l}(n_1,n_2,n_3)\to
a_{\rm l}\left(n_1+\frac{\epsilon}{2}n_2,
n_2+\frac{\epsilon}{2}n_1,\frac{n_3}{1-\epsilon^2/4}\right).
\label{trans1}
\end{equation}
Here $\epsilon$ is a deformation parameter; $n_1$, $n_2$, and $n_3$ are arbitrary integers, and $a_{\rm l}$ is the lattice constant. 

At some critical $\epsilon=\epsilon_{\max}$ the crystal becomes unstable. The breaking stress can be defined as
\begin{equation}
   \sigma_{\max}\equiv\frac{\partial {\cal E}}{\partial \epsilon}{\bigg\vert_{\epsilon=\epsilon_{\max}}},
\label{sig}
\end{equation} 
where ${\cal E}=E_\text{int}/V$ is the internal energy density of the crystal. In our calculations we assume that the crystal is unstable if one or more of its squared phonon frequencies become negative at some wave vector in the first Brillouin zone. The phonon spectrum is calculated in the harmonic-lattice approximation, so that our results are valid for $T \ll T_{\rm m}$, where $T_{\rm m}$ is the melting temperature.

As shown by \cite{KY20}, the plasma screening correction is small and could be neglected. At low temperatures the main contribution to the internal energy comes from the electrostatic (Madelung) energy $U_{\rm M}$. The electrostatic energy of any ordered multi-component Coulomb lattice $U_{\rm M}$ can be written as 
\begin{equation}
U_{\rm M} =N\frac{Z_1^2e^{2}}{a}\zeta~, \label{Mad} 
\end{equation}
where $\zeta$ is called a Madelung constant,
\begin{eqnarray}
\zeta&=&\frac{a}{2N_{\rm cell}}\sum_{lpp'}\frac{{Z}_{p}{Z}_{p'}}{Z^2_1}
  \left(1-\delta_{pp'}\delta_{\textbf{R}_l0}\right) \frac{{\rm erfc}
  \left(A Y_{lpp'}\right)}{Y_{lpp'}} \nonumber \\
  &-&\frac{Aa}{N_{\rm cell}\sqrt{\pi}}\sum_p \frac{Z^2_p}{Z_1^2}
  -\frac{3}{8N_{\rm cell}^2A^2a^2}\sum_{pp'}\frac{Z_{p}Z_{p'}}{Z_1^2} \nonumber \\
  &+&\frac{3}{2N_{\rm cell}^2 a^2}\sum_{mpp'}\frac{{Z}_{p}{Z}_{p'}}{Z_1^2}
    (1-\delta_{\textbf{G}_m0})\frac{1}{G_m^2} \nonumber \\
  &\times& \exp\left[-\frac{G_m^2}{4A^2}+
     i\textbf{G}_m(\boldsymbol{\chi}_p
      -\boldsymbol{\chi}_{p'})\right]~.
\end{eqnarray}
Here ${\textbf{Y}}_{lpp'}={\textbf{R}}_l+\boldsymbol{\chi}_p-\boldsymbol{\chi}_{p'}$, ${\rm erfc}(x)$ is the complementary error function, $\textbf{R}_l$ is a lattice vector, $\boldsymbol{\chi}_{p}$ is the basis vector of the $p$-th ion in the elementary cell  ($p=1\dots N_{\rm cell}$), vectors $\textbf{G}_m$ form a reciprocal lattice, an arbitrary constant $A$ can be chosen as $Aa \approx 2$ (Eq. (\ref{Mad}) does not depend on its specific value). The direction to the nearest neighbor is given by vector $0.5a_{\rm l}(1,1,1)$. Because our crystal is strongly ordered, the charge number of an ion is determined by its position in the elementary cell (see, \citealt{KB15}). For the binary bcc lattice the Madelung constant depends only on $\alpha\equiv Z_2/Z_1$, and we can assume that $\alpha \geq 1$.

Eq. (\ref{Mad}) is universal and valid for any  multi-component crystal including a deformed one. As shown by \cite{BK17,BC18}, the properties of Coulomb crystals depend on the direction of deformation. However, in the neutron star crust the orientation of the crystal is unknown, and for solving astrophysical problems one can chose any arbitrary orientation. We will consider only a few deformations of a binary bcc lattice, which conserves the volume of the elementary cell.

The binary bcc lattice is unstable at $\alpha>3.596$ without any deformation (e.g., \citealt{K20}). For cases under consideration, the mass of the ion affects neither the stability nor energy and, hence, it does not affect the breaking stress.

\begin{figure}
\includegraphics[width=0.45\textwidth]{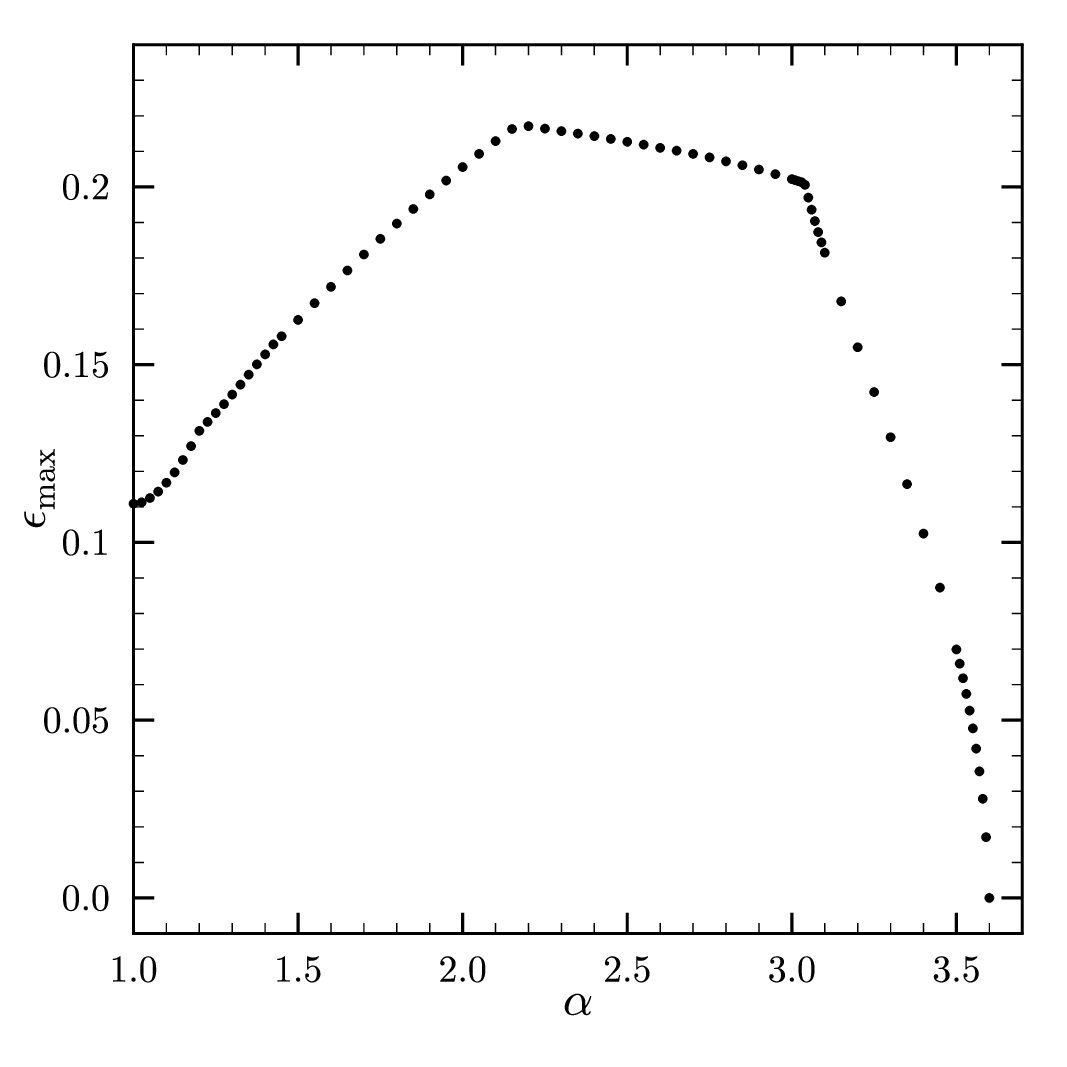}
\hspace{5mm}
\caption{The maximum allowable value of the deformation parameter $\epsilon$ as a function of $\alpha$ for the deformation described by Eq. (\ref{trans1}).}
\label{f1}
\end{figure}
For a given direction of deformation and $\alpha$, we calculate $\epsilon_{\rm max}$ and obtain the breaking stress $\sigma_{\rm max}$ from Eq. (\ref{sig}). First, let us to consider a shear deformation described by Eq. (\ref{trans1}). For this deformation, in Fig. \ref{f1} we plot $\epsilon_{\rm max}$ for selected values of $\alpha$. For different $\alpha$, complex-valued modes appear in different places of the first Brillouin zone. Therefore, the dependence of $\epsilon_{\max}$ on $\alpha$ is non-monotonic. For a crystal with $\alpha=1$ one has $\epsilon_{\rm max}=0.1109$. $\epsilon_{\max}$ reaches maximum 0.217 at $\alpha=2.2$ and rapidly decreases to zero at $\alpha>3.1$.

\begin{figure}
\includegraphics[width=0.45\textwidth]{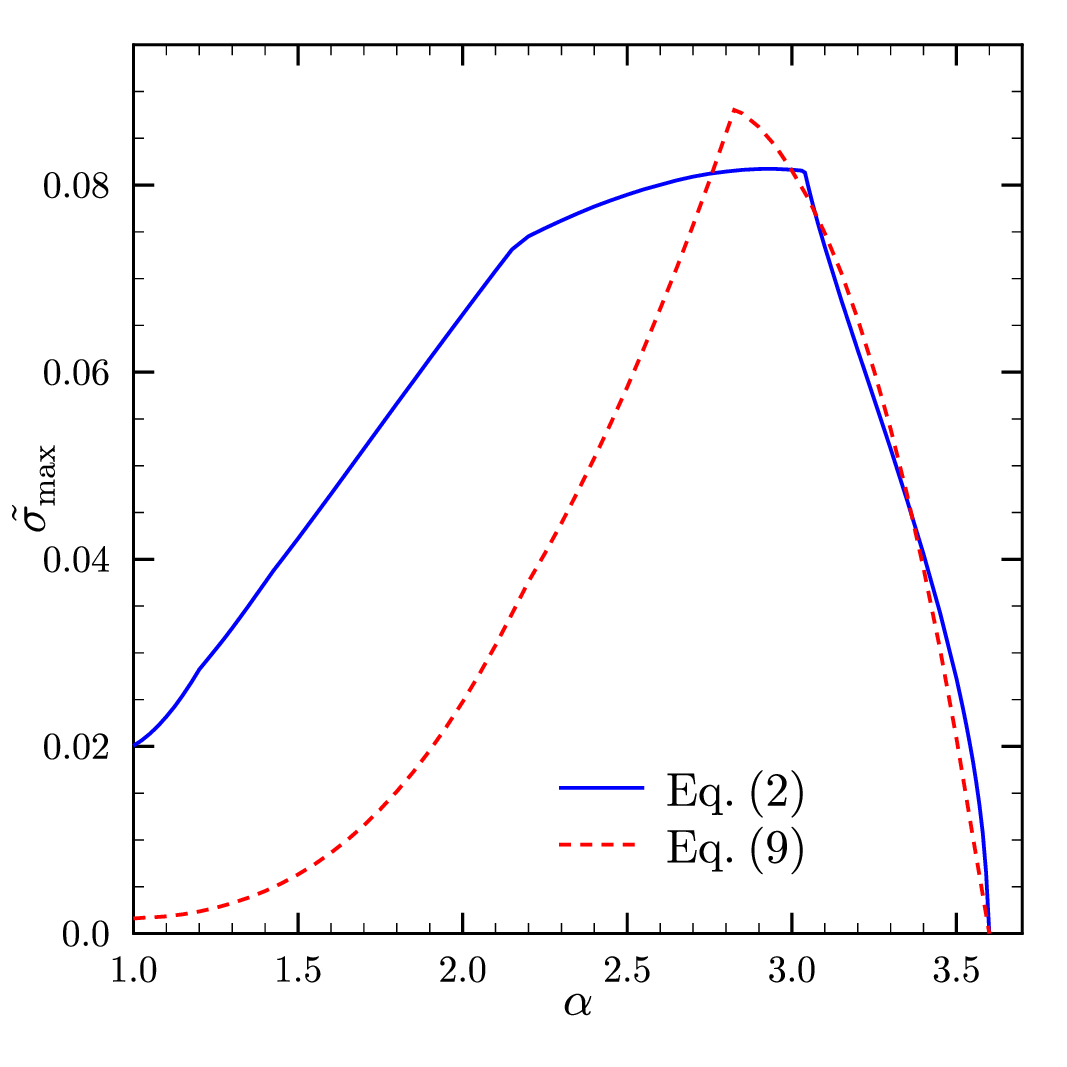}
\hspace{5mm}
\caption{The dependence of $\sigma_{\max}$ on $\alpha$ for deformations described by Eq. (\ref{trans1}) and Eq. (\ref{trans2}).}
\label{f2}
\end{figure}
Fig. \ref{f2} shows the dependence of $\widetilde{\sigma}$ on $\alpha$, where 
\begin{equation}
   \sigma_{\max}=n\frac{Z_1^2 e^2}{a}{\widetilde{\sigma}}~, \qquad \widetilde{\sigma}=\frac{\partial {\zeta}}{\partial \epsilon}{\bigg\vert_{\epsilon=\epsilon_{\max}}}.
\label{sig2}
\end{equation} 
The solid blue line in Fig. \ref{f2} plots $\sigma_{\max}$ versus $\alpha$ for the deformation described by Eq. (\ref{trans1}). For the one-component lattice we obtain $\widetilde{\sigma}=0.02007$, which agrees with previous calculations by \cite{KY20} and with the results reported by \citet{CH10}. The maximum of $\widetilde{\sigma}$ is 0.0817 at $\alpha=2.95$, and it is more than four times larger than at $\alpha=1$.

For $\alpha < 1.75$ the dependence of $\widetilde{\sigma}$ on $\alpha$ can be fitted (with a few percent error) as
\begin{equation}
   {\widetilde{\sigma}}=-0.0279+0.0467\alpha.
\label{apr1}
\end{equation}

\begin{figure}
\includegraphics[width=0.45\textwidth]{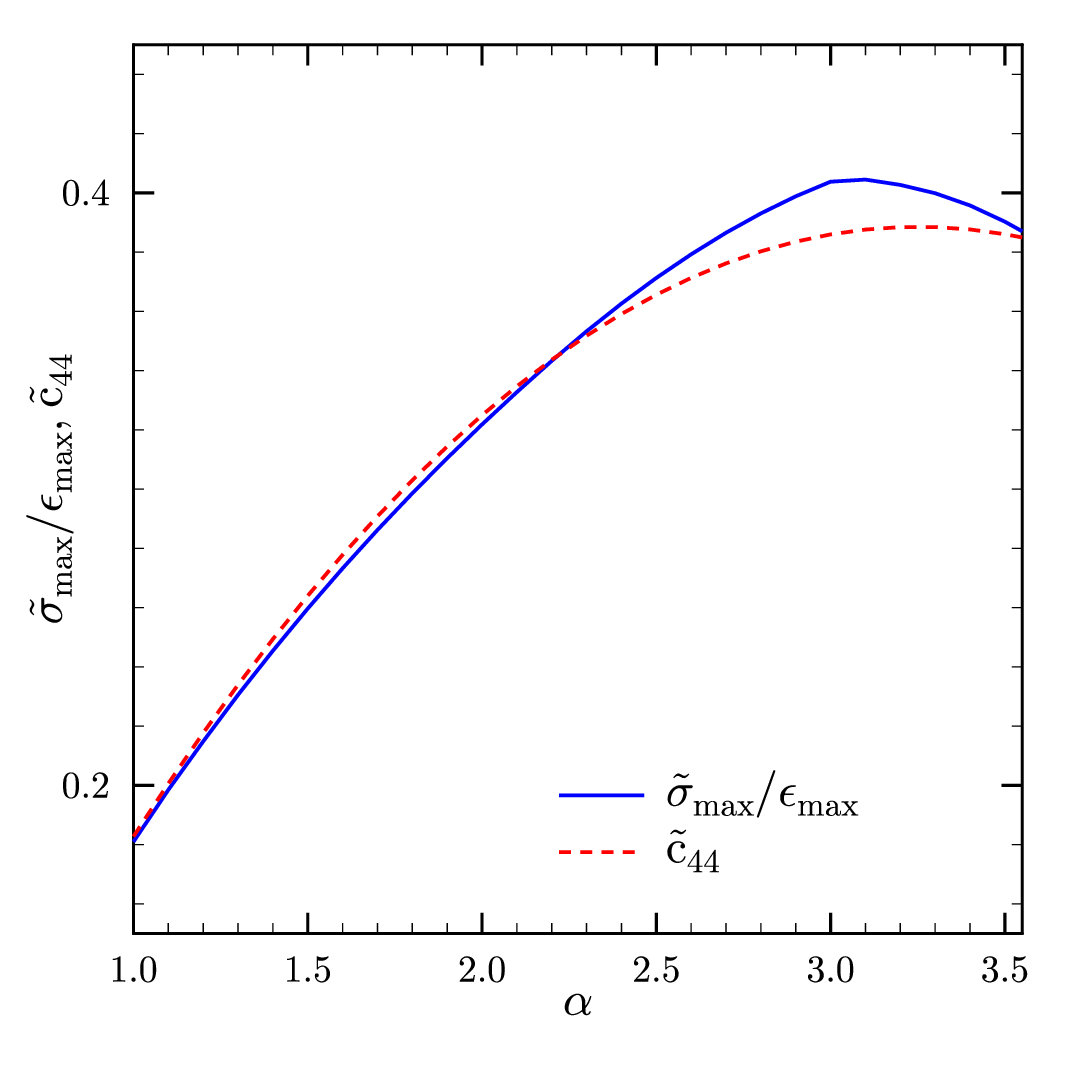}
\hspace{5mm}
\caption{The dependence of $\widetilde{\sigma}_{\max}/\epsilon_{\max}$ and $\widetilde{c}_{44}$ on $\alpha$ for deformation described by Eq. (\ref{trans1}).}
\label{f3}
\end{figure}
According to the traditional assumption, the breaking stress is  
\begin{equation}
    \sigma_{\max}^{\rm trad} \approx \epsilon_{\max}\mu,
\label{trad}    
\end{equation}
where $\mu$ is the effective shear modulus (see, e.g., \citealt{BC18} for details). Our method of calculating $\sigma_{\max}$ allows us to check Eq. (\ref{trad}). In Eq. (\ref{trad}) one uses averaged values of $\epsilon_{\max}$ and $\mu$. Since we calculated $\epsilon_{\max}$ for one specific 
shear deformation, in Eq. (\ref{trad}) it is more accurate to use elastic coefficient $c_{44}$ instead of $\mu$. The elastic properties of binary bcc lattice were studied by \cite{K19}, who obtained $c_{44}=nZ_1^2e^2\widetilde{c}_{44}/a$ and $\widetilde{c}_{44}=-0.040583(1+\alpha^2)+0.263936\alpha$. In Fig. \ref{f3} we compare $\widetilde{\sigma}_{\max}/\epsilon_{\max}$ (the solid blue line) and $\widetilde{c}_{44}$ (the dashed red line) in a range of $\alpha$. We see that for the chosen deformation the traditional assumption works well enough. The difference does not exceed a few percent.

We have also considered another type of deformation of the binary bcc lattice,
\begin{equation}
a_{\rm l}(n_1,n_2,n_3)\to a_{\rm l} \left(\epsilon n_1,n_2/\epsilon,n_3\right),
\label{trans2}
\end{equation}
where $\epsilon > 1$ is again the deformation parameter. 
We have calculated the breaking stress $\sigma_{\rm max}$ in the same way as before. The results are presented in Fig. \ref{f2} by the red dashed line. $\widetilde{\sigma}$ monotonically increases from 0.00168 at $\alpha=1$ to 0.088 at $\alpha=2.825$, while $\epsilon_{\max}$ does not exceed 1.08 at any possible $\alpha$. Since the binary bcc lattices with $\alpha>3.596$ are unstable, we expect that at $\alpha$ slightly less than 3.596 the breaking properties do not depend on the direction of deformation. This agrees with Fig. \ref{f2}, where for both deformations under consideration the dependencies of $\sigma_{\max}$ on $\alpha$ at $\alpha>3$ are similar.

Deformation described by Eq. (\ref{trans2}) stretch the lattice, so it should be associated with the elastic modulus $c_{11}-c_{12}$. For this deformation  Eq. (\ref{trad}) becomes invalid. The difference between $\sigma$ and $\sigma_{\max}^{\rm trad}$ can exceed a factor of several.

Other deformations show similar dependence of $\sigma_{\max}$ on $\alpha$.

\section{Discussion and conclusions}   
We have studied deformed Coulomb crystals of atomic nuclei and analyzed breaking properties of the neutron star crust. It is shown that, if the crust consist of two types of ions at any fixed density, its breaking stress could be noticeably greater than those typical for one-component crystal. For all cases under consideration, $\widetilde{\sigma}$ increases with $\alpha$ up to $0.06$-$0.09$ at $\alpha\approx2.5$, while at higher $\alpha$ it decreases and reaches zero at $\alpha\approx3.596$. For the same directions of deformations the critical stain does not increase that much but it can vary nonmonotonically.

Previously, the breaking properties of multi-component solids in the neutron star crust were studied by \cite{HK09} and \cite{HH12}. In the first paper a complex and heterogeneous system of ions was considered. Therefore, accurate comparison of their and our results is difficult. 
Among all mixtures considered by \cite{HH12} the mixture of $^{54}$Fe, $^{56}$Fe and $^{58}$Ni is of the great interest because fractions of iron and nickel are approximately of the same order of magnitude. For this mixture $\widetilde{\sigma} \approx 0.02$ is obtained to be slightly greater than for a pure iron crystal (see fig 2 in \cite{HH12}). This difference is in good agreement with our calculations, where $\widetilde{\sigma}$ increases at small $\alpha$ for any deformation.

It is instructive to describe composition of the neutron star envelopes by the effective impurity parameter $Q$ \citep[e.g.,][]{HPY2007} and the average charge number $\langle Z\rangle$. For a binary bcc lattice one has $Q=0.25(Z_2-Z_1)^2$ and $\langle Z\rangle=0.5(Z_2+Z_1)$. In this notation Eq. (\ref{apr1}) could be written as
\begin{eqnarray}   
  \sigma_{\max}&=&n\frac{\langle Z\rangle^2e^2}{a} \hat{\sigma}~, \label{apr2} \\
   \hat{\sigma}&=&0.0189+0.0558\frac{\sqrt{Q}}{\langle Z\rangle^2}-0.0746\frac{Q}{\langle Z\rangle^2}~. \nonumber
\end{eqnarray}
For the one-component crystal $\hat{\sigma}$ takes its standard value 0.0189. At $\langle Z\rangle=19$ and $Q=3$ Eq. (\ref{apr2}) gives $\hat{\sigma}=0.0258$; at $\langle Z\rangle=22$ and $Q=3$ the changes are small, $\hat{\sigma}=0.0251$; while at $\langle Z\rangle=22$ and $Q=5$ we have  $\hat{\sigma}=0.0277$. So a small change for $Q$ can lead to a larger change in $\sigma_{\max}$. 

Some models suggest that the crust is highly heterogeneous. For instance, one could have $\langle Z\rangle=30$ and $Q=40$, but such combination is impossible for a binary crystal, and therefore Eq.(\ref{apr2}) is useless. Note that Eq. (\ref{apr2}) is constructed for an idealized system of a strictly ordered binary crystal without impurities. Hence Eq. (\ref{apr2}) can be treated as an upper limit for realistic mixtures.

Our results can be useful to simulate various processes involving deformed Coulomb crystals in neutron stars, such as evolution of the magnetic fields, quasi-periodic oscillations and pulsar glitches \citep[e.g.,][]{UCB00,VRP13,BL14,GTWZ20,ZG22}.

\section*{Acknowledgments}
The author is deeply grateful to D.~G. Yakovlev for help and discussions.
The work was supported by the Russian
Science Foundation (grant 19-12-00133). 

\section*{Data availability}
The data underlying this article are available upon request to the author.

\bibliographystyle{mnras}

\bibliography{BibList}

\begin{thebibliography}{}
\makeatletter
\relax
\def\mn@urlcharsother{\let\do\@makeother \do\$\do\&\do\#\do\^\do\_\do\%\do\~}
\def\mn@doi{\begingroup\mn@urlcharsother \@ifnextchar [ {\mn@doi@}
  {\mn@doi@[]}}
\def\mn@doi@[#1]#2{\def\@tempa{#1}\ifx\@tempa\@empty \href
  {http://dx.doi.org/#2} {doi:#2}\else \href {http://dx.doi.org/#2} {#1}\fi
  \endgroup}
\def\mn@eprint#1#2{\mn@eprint@#1:#2::\@nil}
\def\mn@eprint@arXiv#1{\href {http://arxiv.org/abs/#1} {{\tt arXiv:#1}}}
\def\mn@eprint@dblp#1{\href {http://dblp.uni-trier.de/rec/bibtex/#1.xml}
  {dblp:#1}}
\def\mn@eprint@#1:#2:#3:#4\@nil{\def\@tempa {#1}\def\@tempb {#2}\def\@tempc
  {#3}\ifx \@tempc \@empty \let \@tempc \@tempb \let \@tempb \@tempa \fi \ifx
  \@tempb \@empty \def\@tempb {arXiv}\fi \@ifundefined
  {mn@eprint@\@tempb}{\@tempb:\@tempc}{\expandafter \expandafter \csname
  mn@eprint@\@tempb\endcsname \expandafter{\@tempc}}}

\bibitem[\protect\citeauthoryear{{Abbott} et~al.,}{{Abbott} et~al.}{2019}]{A19}
{Abbott} B.~P.,  et~al., 2019, \mn@doi [\apj] {10.3847/1538-4357/ab113b}, \href
  {https://ui.adsabs.harvard.edu/abs/2019ApJ...875..122A} {875, 122}

\bibitem[\protect\citeauthoryear{{Baiko}}{{Baiko}}{2022}]{B22}
{Baiko} D.~A.,  2022, \mn@doi [\mnras] {10.1093/mnras/stac2693}, \href
  {https://ui.adsabs.harvard.edu/abs/2022MNRAS.517.3962B} {517, 3962}

\bibitem[\protect\citeauthoryear{{Baiko} \& {Chugunov}}{{Baiko} \&
  {Chugunov}}{2018}]{BC18}
{Baiko} D.~A.,  {Chugunov} A.~I.,  2018, \mn@doi [\mnras]
  {10.1093/mnras/sty2259}, 480, 5511

\bibitem[\protect\citeauthoryear{{Baiko} \& {Kozhberov}}{{Baiko} \&
  {Kozhberov}}{2017}]{BK17}
{Baiko} D.~A.,  {Kozhberov} A.~A.,  2017, \mn@doi [\mnras]
  {10.1093/mnras/stx1270}, 470, 517

\bibitem[\protect\citeauthoryear{{Beloborodov} \& {Levin}}{{Beloborodov} \&
  {Levin}}{2014}]{BL14}
{Beloborodov} A.~M.,  {Levin} Y.,  2014, \mn@doi [\apjl]
  {10.1088/2041-8205/794/2/L24}, 794, L24

\bibitem[\protect\citeauthoryear{{Bera}, {Jones}  \& {Andersson}}{{Bera}
  et~al.}{2020}]{BJA20}
{Bera} P.,  {Jones} D.~I.,   {Andersson} N.,  2020, \mn@doi [\mnras]
  {10.1093/mnras/staa3015}, \href
  {https://ui.adsabs.harvard.edu/abs/2020MNRAS.499.2636B} {499, 2636}

\bibitem[\protect\citeauthoryear{{Chamel} \& {Haensel}}{{Chamel} \&
  {Haensel}}{2008}]{CH08}
{Chamel} N.,  {Haensel} P.,  2008, \mn@doi [Living Reviews in Relativity]
  {10.12942/lrr-2008-10}, 11, 10

\bibitem[\protect\citeauthoryear{{Chugunov} \& {Horowitz}}{{Chugunov} \&
  {Horowitz}}{2010}]{CH10}
{Chugunov} A.~I.,  {Horowitz} C.~J.,  2010, \mn@doi [\mnras]
  {10.1111/j.1745-3933.2010.00903.x}, 407, L54

\bibitem[\protect\citeauthoryear{{Chugunov} \& {Horowitz}}{{Chugunov} \&
  {Horowitz}}{2012}]{CH12}
{Chugunov} A.~I.,  {Horowitz} C.~J.,  2012, \mn@doi [Contributions to Plasma
  Physics] {10.1002/ctpp.201100075}, 52, 122

\bibitem[\protect\citeauthoryear{{Daligault} \& {Gupta}}{{Daligault} \&
  {Gupta}}{2009}]{DG09}
{Daligault} J.,  {Gupta} S.,  2009, \mn@doi [\apj]
  {10.1088/0004-637X/703/1/994}, \href
  {https://ui.adsabs.harvard.edu/abs/2009ApJ...703..994D} {703, 994}

\bibitem[\protect\citeauthoryear{{De Grandis}, {Turolla}, {Wood}, {Zane},
  {Taverna}  \& {Gourgouliatos}}{{De Grandis} et~al.}{2020}]{GTWZ20}
{De Grandis} D.,  {Turolla} R.,  {Wood} T.~S.,  {Zane} S.,  {Taverna} R.,
  {Gourgouliatos} K.~N.,  2020, \mn@doi [\apj] {10.3847/1538-4357/abb6f9},
  \href {https://ui.adsabs.harvard.edu/abs/2020ApJ...903...40D} {903, 40}

\bibitem[\protect\citeauthoryear{{Fattoyev}, {Horowitz}  \& {Lu}}{{Fattoyev}
  et~al.}{2018}]{FHL18}
{Fattoyev} F.~J.,  {Horowitz} C.~J.,   {Lu} H.,  2018, arXiv e-prints, p.
  arXiv:1804.04952

\bibitem[\protect\citeauthoryear{{Haensel}, {Potekhin}  \&
  {Yakovlev}}{{Haensel} et~al.}{2007}]{HPY2007}
{Haensel} P.,  {Potekhin} A.~Y.,   {Yakovlev} D.~G.,  2007, {Neutron Stars. 1.
  Equation of State and Structure}.
Springer, New York

\bibitem[\protect\citeauthoryear{{Hoffman} \& {Heyl}}{{Hoffman} \&
  {Heyl}}{2012}]{HH12}
{Hoffman} K.,  {Heyl} J.,  2012, \mn@doi [\mnras]
  {10.1111/j.1365-2966.2012.21921.x}, 426, 2404

\bibitem[\protect\citeauthoryear{{Horowitz} \& {Berry}}{{Horowitz} \&
  {Berry}}{2009}]{HB09}
{Horowitz} C.~J.,  {Berry} D.~K.,  2009, \mn@doi [\prc]
  {10.1103/PhysRevC.79.065803}, \href
  {https://ui.adsabs.harvard.edu/abs/2009PhRvC..79f5803H} {79, 065803}

\bibitem[\protect\citeauthoryear{{Horowitz} \& {Kadau}}{{Horowitz} \&
  {Kadau}}{2009}]{HK09}
{Horowitz} C.~J.,  {Kadau} K.,  2009, \mn@doi [\prl]
  {10.1103/PhysRevLett.102.191102}, 102, 191102

\bibitem[\protect\citeauthoryear{Jaranowski, Kr\'olak  \& Schutz}{Jaranowski
  et~al.}{1998}]{JKS98}
Jaranowski P.,  Kr\'olak A.,   Schutz B.~F.,  1998, \mn@doi [Phys. Rev. D]
  {10.1103/PhysRevD.58.063001}, 58, 063001

\bibitem[\protect\citeauthoryear{{Johnson-McDaniel} \&
  {Owen}}{{Johnson-McDaniel} \& {Owen}}{2013}]{JO13}
{Johnson-McDaniel} N.~K.,  {Owen} B.~J.,  2013, \mn@doi [\prd]
  {10.1103/PhysRevD.88.044004}, \href
  {https://ui.adsabs.harvard.edu/abs/2013PhRvD..88d4004J} {88, 044004}

\bibitem[\protect\citeauthoryear{{Kaspi} \& {Beloborodov}}{{Kaspi} \&
  {Beloborodov}}{2017}]{KB17}
{Kaspi} V.~M.,  {Beloborodov} A.~M.,  2017, \mn@doi [\araa]
  {10.1146/annurev-astro-081915-023329}, \href
  {https://ui.adsabs.harvard.edu/abs/2017ARA&A..55..261K} {55, 261}

\bibitem[\protect\citeauthoryear{{Kerin} \& {Melatos}}{{Kerin} \&
  {Melatos}}{2022}]{KM22}
{Kerin} A.~D.,  {Melatos} A.,  2022, \mn@doi [\mnras] {10.1093/mnras/stac1351},
  \href {https://ui.adsabs.harvard.edu/abs/2022MNRAS.514.1628K} {514, 1628}

\bibitem[\protect\citeauthoryear{{Kozhberov}}{{Kozhberov}}{2019}]{K19}
{Kozhberov} A.~A.,  2019, \mn@doi [\mnras] {10.1093/mnras/stz1151}, \href
  {https://ui.adsabs.harvard.edu/abs/2019MNRAS.486.4473K} {486, 4473}

\bibitem[\protect\citeauthoryear{{Kozhberov}}{{Kozhberov}}{2020}]{K20}
{Kozhberov} A.~A.,  2020, \mn@doi [Contributions to Plasma Physics]
  {10.1002/ctpp.202000021}, 60, e202000021

\bibitem[\protect\citeauthoryear{{Kozhberov} \& {Baiko}}{{Kozhberov} \&
  {Baiko}}{2015}]{KB15}
{Kozhberov} A.~A.,  {Baiko} D.~A.,  2015, \mn@doi [Physics of Plasmas]
  {10.1063/1.4930215}, \href
  {https://ui.adsabs.harvard.edu/abs/2015PhPl...22i2903K} {22, 092903}

\bibitem[\protect\citeauthoryear{{Kozhberov} \& {Yakovlev}}{{Kozhberov} \&
  {Yakovlev}}{2020}]{KY20}
{Kozhberov} A.~A.,  {Yakovlev} D.~G.,  2020, \mn@doi [\mnras]
  {10.1093/mnras/staa2715}, \href
  {https://ui.adsabs.harvard.edu/abs/2020MNRAS.498.5149K} {498, 5149}

\bibitem[\protect\citeauthoryear{{Kuan}, {Suvorov}  \& {Kokkotas}}{{Kuan}
  et~al.}{2021}]{KSK21}
{Kuan} H.-J.,  {Suvorov} A.~G.,   {Kokkotas} K.~D.,  2021, \mn@doi [\mnras]
  {10.1093/mnras/stab2658}, \href
  {https://ui.adsabs.harvard.edu/abs/2021MNRAS.508.1732K} {508, 1732}

\bibitem[\protect\citeauthoryear{{Lander}}{{Lander}}{2016}]{L16}
{Lander} S.~K.,  2016, \mn@doi [\apjl] {10.3847/2041-8205/824/2/L21}, 824, L21

\bibitem[\protect\citeauthoryear{{Lander} \& {Gourgouliatos}}{{Lander} \&
  {Gourgouliatos}}{2019}]{LG19}
{Lander} S.~K.,  {Gourgouliatos} K.~N.,  2019, \mn@doi [\mnras]
  {10.1093/mnras/stz1042}, \href
  {https://ui.adsabs.harvard.edu/abs/2019MNRAS.486.4130L} {486, 4130}

\bibitem[\protect\citeauthoryear{{Lander}, {Andersson}, {Antonopoulou}  \&
  {Watts}}{{Lander} et~al.}{2015}]{LAAW15}
{Lander} S.~K.,  {Andersson} N.,  {Antonopoulou} D.,   {Watts} A.~L.,  2015,
  \mn@doi [\mnras] {10.1093/mnras/stv432}, \href
  {https://ui.adsabs.harvard.edu/abs/2015MNRAS.449.2047L} {449, 2047}

\bibitem[\protect\citeauthoryear{{Li}, {Levin}  \& {Beloborodov}}{{Li}
  et~al.}{2016}]{LL16}
{Li} X.,  {Levin} Y.,   {Beloborodov} A.~M.,  2016, \mn@doi [\apj]
  {10.3847/1538-4357/833/2/189}, 833, 189

\bibitem[\protect\citeauthoryear{{Owen}}{{Owen}}{2005}]{O05}
{Owen} B.~J.,  2005, \mn@doi [\prl] {10.1103/PhysRevLett.95.211101}, 95, 211101

\bibitem[\protect\citeauthoryear{{Passamonti}, {Andersson}  \&
  {Pnigouras}}{{Passamonti} et~al.}{2021}]{PAP21}
{Passamonti} A.,  {Andersson} N.,   {Pnigouras} P.,  2021, \mn@doi [\mnras]
  {10.1093/mnras/stab870}, \href
  {https://ui.adsabs.harvard.edu/abs/2021MNRAS.504.1273P} {504, 1273}

\bibitem[\protect\citeauthoryear{{Perna} \& {Pons}}{{Perna} \&
  {Pons}}{2011}]{PP11}
{Perna} R.,  {Pons} J.~A.,  2011, \mn@doi [\apjl]
  {10.1088/2041-8205/727/2/L51}, \href
  {https://ui.adsabs.harvard.edu/abs/2011ApJ...727L..51P} {727, L51}

\bibitem[\protect\citeauthoryear{{Piekarewicz}, {Fattoyev}  \&
  {Horowitz}}{{Piekarewicz} et~al.}{2014}]{PFH14}
{Piekarewicz} J.,  {Fattoyev} F.~J.,   {Horowitz} C.~J.,  2014, \mn@doi [\prc]
  {10.1103/PhysRevC.90.015803}, 90, 015803

\bibitem[\protect\citeauthoryear{{Rencoret}, {Aguilera-G{\'o}mez}  \&
  {Reisenegger}}{{Rencoret} et~al.}{2021}]{RAR21}
{Rencoret} J.~A.,  {Aguilera-G{\'o}mez} C.,   {Reisenegger} A.,  2021, \mn@doi
  [\aap] {10.1051/0004-6361/202141499}, \href
  {https://ui.adsabs.harvard.edu/abs/2021A&A...654A..47R} {654, A47}

\bibitem[\protect\citeauthoryear{{Shapiro} \& {Teukolsky}}{{Shapiro} \&
  {Teukolsky}}{1983}]{ST1983}
{Shapiro} S.~L.,  {Teukolsky} S.~A.,  1983, {Black holes, white dwarfs, and
  neutron stars: The physics of compact objects}.
Wiley-Interscience, New York

\bibitem[\protect\citeauthoryear{{Shchechilin} \& {Chugunov}}{{Shchechilin} \&
  {Chugunov}}{2019}]{SC19}
{Shchechilin} N.~N.,  {Chugunov} A.~I.,  2019, \mn@doi [\mnras]
  {10.1093/mnras/stz2838}, \href
  {https://ui.adsabs.harvard.edu/abs/2019MNRAS.490.3454S} {490, 3454}

\bibitem[\protect\citeauthoryear{{Thompson}, {Yang}  \& {Ortiz}}{{Thompson}
  et~al.}{2017}]{THN17}
{Thompson} C.,  {Yang} H.,   {Ortiz} N.,  2017, \mn@doi [\apj]
  {10.3847/1538-4357/aa6c30}, \href
  {https://ui.adsabs.harvard.edu/abs/2017ApJ...841...54T} {841, 54}

\bibitem[\protect\citeauthoryear{{Ushomirsky}, {Cutler}  \&
  {Bildsten}}{{Ushomirsky} et~al.}{2000}]{UCB00}
{Ushomirsky} G.,  {Cutler} C.,   {Bildsten} L.,  2000, \mn@doi [\mnras]
  {10.1046/j.1365-8711.2000.03938.x}, \href
  {https://ui.adsabs.harvard.edu/abs/2000MNRAS.319..902U} {319, 902}

\bibitem[\protect\citeauthoryear{{Vigan{\`o}}, {Rea}, {Pons}, {Perna},
  {Aguilera}  \& {Miralles}}{{Vigan{\`o}} et~al.}{2013}]{VRP13}
{Vigan{\`o}} D.,  {Rea} N.,  {Pons} J.~A.,  {Perna} R.,  {Aguilera} D.~N.,
  {Miralles} J.~A.,  2013, \mn@doi [\mnras] {10.1093/mnras/stt1008}, \href
  {https://ui.adsabs.harvard.edu/abs/2013MNRAS.434..123V} {434, 123}

\bibitem[\protect\citeauthoryear{{Zhou}, {G{\"u}gercino{\u{g}}lu}, {Yuan}, {Ge}
   \& {Yu}}{{Zhou} et~al.}{2022}]{ZG22}
{Zhou} S.,  {G{\"u}gercino{\u{g}}lu} E.,  {Yuan} J.,  {Ge} M.,   {Yu} C.,
  2022, \mn@doi [Universe] {10.3390/universe8120641}, \href
  {https://ui.adsabs.harvard.edu/abs/2022Univ....8..641Z} {8, 641}

\makeatother
\end{thebibliography}

\end{document}